\documentstyle[12pt,epsf]{article}
\newcommand{\sect}[1]{\setcounter{equation}{0}\section{#1}}

\newcommand{\NP}[1]{Nucl. Phys.\ {\bf #1}\ }

\newcommand{\PL}[1]{Phys. Lett.\ {\bf #1}\ }
\newcommand{\PR}[1]{Phys. Rev.\ {\bf #1}\ }

\newcommand{\ZP}[1]{Z. Phys. \ {\bf #1}\ }

\def\beg{\begin{equation}}
\def\begar{\begin{eqnarray}}
\def\ee{\end{equation}}
\def\ea{\end{eqnarray}}

\newcommand{\pref}[1]{(\ref{#1})}                
\newcommand{\plabel}[1]{\label{#1}}              
\newcommand{\pcite}[1]{\cite{#1}}                
\newcommand{\pbib}[1]{\bibitem{#1}}              

\def\2{{1\over2}} \def\4{{1\over4}} \def\52{{5\over2}} \def\6{\partial }
\def\({\left(} \def\){\right)} \def\<{\langle } \def\>{\rangle }

\def\nn{\nonumber}

\def \as{\alpha_{\rm s}}
\let\a=\alpha  \let\g=\gamma \let\d=\delta
   
   \let\l=\lambda
\let\m=\mu   
\def\kv{{\bf k}}\def\pv{{\bf p}} \def\qv{{\bf q}} \def\tv{{\bf t}} 
\def\sv{{\bf s}} 
\def\ggv{{\bf \gamma}}
   \let\o=\omega

\def\CV{{\cal V}}

\newcommand{\ftt}[2]{{d^{#1}{#2}\over (2\pi)^{#1}}}

\begin{document}
\begin{titlepage}
\begin{flushright}
CERN-TH/96-27\\
TUW-96-02\\
\end{flushright}

\vspace{1cm}
\begin{center}
\begin{Large}
{\bf QCD Box Graphs and the 
Quark-Antiquark Potential}\\[2cm]
\end{Large} 
{\large W. Kummer
\footnote{wkummer@tph.tuwien.ac.at}, W. M\"odritsch}
\footnote{wmoedrit@ecxph.tuwien.ac.at}\\
{\it Institut f\"ur Theoretische Physik, Techn. Univ. Wien\\
Wiedner Hauptstra\ss e 8-10, A-1040 Vienna, Austria}\\[1cm]
{\large A. Vairo}
\footnote{vairo@vxcern.cern.ch}\\
{\it Theory Division, CERN, CH-1211 Geneva 23, Switzerland,\\
INFN, Sezione di Bologna, Via Irnerio 46, 40126 Bologna, Italy}
\end{center}

\vspace{1cm}
\begin{abstract}
{The $t\bar{t}$ system allows a truly perturbative treatment of
the potential. Completing previous computations, we calculate the 
contributions of QCD box graph corrections, which make the  
``relativistic'' $O(\a^4)$ corrections in the non-Abelian case differ 
from the well known  corrections of the same order in QED. 
\vspace{0.5cm}
}
\end{abstract}

\vspace{1.5cm}
\par\noindent 
CERN-TH/96-27\\ 
TUW-96-02\\   
February 1996\\
\end{titlepage}

\vfill

\setcounter{footnote}{0} 

\sect{Introduction}

The literature on the perturbative treatment of the quark-antiquark 
potential is very abundant.  But even for the bottom quark these calculations
suffer from the drawback that non-perturbative contributions can be estimated
to be as large as the leading Coulomb part, and thus that their perturbative
inclusion is not justified. On the other hand, phenomenological potentials 
with a large number of free parameters are used extensively \pcite{Gara}.
The situation is completely different for the top quark. 
With its mass of $\approx 180$ GeV it is heavy enough to allow
a completely perturbative treatment. The drawback of this heavy mass is that, 
owing to its weak decay, individual bound state levels are not directly
observable. However, this turns into a virtue since this large decay width
effectively cuts off non-perturbative contributions in the calculation of the 
cross sections near threshold \pcite{Khoze}. For computations in this region 
the input of a perturbative QCD potential is justified and it is thus 
desirable to investigate to which extent we are able to calculate such a 
potential. Unfortunately, despite the large amount of literature, the number 
of studies including reliable QCD perturbative calculations beyond the 
lowest orders is actually small. For example in most cases higher order 
corrections are derived from on shell scattering amplitudes \pcite{scatt}. 
This is not a consistent approach since different powers 
give contributions to the same order in the energy shift of bound states. 
The present article is also intended to shed some light on the calculation 
of spin-dependent forces up to $O(\as^5 \ln \as)$ where perturbation 
theory is likely to be applicable even for lighter quarks. 
 
Recently a consistent treatment on the basis of the Bethe-Salpeter (BS) 
equation was given \pcite{KM1} where the level shifts themselves have 
been used as ordering parameter. There it was emphasized that it 
would be incorrect to extrapolate from the absence of corrections 
to $O(\as^4)$ other than the tree graphs in the QED case 
to QCD because gluon splitting allows new types of graphs. 
Therefore we will discuss here possible sources of new corrections to 
that particular $O(\as^4)$ in some detail.

\sect{QCD two-loop box graphs}

The two-loop box graphs on which we concentrate in this letter are 
represented in fig. \ref{dfig5}. As we adopt the Coulomb gauge, the 
full gluon propagator (graphically represented by a wavy line) 
can be split, as usual, into a Coulomb part (broken line) and a transverse 
part (curly line). Contributions to the two-loop 
box graphs, which do not appear in fig. \ref{dfig5}, may be checked to give 
higher-order corrections.  

In \pcite{KM1} an exact zeroth-order solution has been derived for the  
toponium BS equation, generalizing the Barbieri-Remiddi equation \pcite{BR}
to the case of unstable particles. For our purposes it is sufficient 
to consider the corresponding zeroth-order wave functions $\chi$ and 
$\bar{\chi}$ in a non-relativistic  approximation:
\begar
\chi(p)^{\rm nr} &=& 
\frac{\sqrt{2} i \tilde{\o}_n}{p_0^2-\tilde{\o}_n^2} \phi_{nlm}(\pv) \, S
= -\g_0 \bar{\chi}(p)^{\rm nr} \g_0  \, ,
\plabel{xapp}
\ea
where $\tilde{\o}_n := (\pv^2/m - E_n)/2 - i\epsilon$, 
$E_n = -m \a^2/4 n^2$ are the Bohr levels, 
$\a := (4/3)g^2/4\pi = (4/3)\as$, $\phi_{nlm}$ 
the Schr\"odinger-Coulomb wave functions, $\chi$ and $\bar{\chi}$ are written as
spinor matrices with $S$ a $4\times4$ matrix representing 
the spin state of the particle-antiparticle system: 
\begar
S &=& \left\{ 
\begin{array}{l@{\quad:\quad}l}
                \g_5 \l^-               &  \mbox{singlet} \\
                {\bf a}_m {\bf \g} \l^- &  \mbox{triplet} \quad (m=0,\pm1 
                                     \quad {\bf a}_m^2 = 1)
\end{array}  \right.  \nn \\
\l^{\pm} &:=&  {1\pm \g_0 \over 2} \, .
\plabel{spin}
\ea
All the following contributions are evaluated between $S$ and 
$-\gamma_0 S \gamma_0$ (times a normalization factor $1/6$).

First we will consider the QCD box graph of fig. \ref{dfig5}a. 
Let $p_1$, $-p_2$, be the incoming and $p_1^\prime$, $-p_2^\prime$,  
the outgoing momenta of the particle and antiparticle. 
We define the total momentum $P := p_1-p_2 = p_1^\prime-p_2^\prime$, 
the relative momentum of the incoming particles $p := (p_1+p_2)/2$ and 
$q := p_2^\prime - p_2 = p_1^\prime - p_1$.
Our present correction is put between BS wave functions, in agreement
with the general approach of BS perturbation theory. 
In that formalism, after performing the colour and spinor indices traces, 
the perturbation kernel of graph \ref{dfig5}a can be written as 
\footnote{All calculations have been  performed with the help 
of the program of symbolic manipulations FORM \cite{Ver}.}
\begar
-i H_{\ref{dfig5}.a} &=&12 \,i\,g^6 \int \ftt{4}{t}\ftt{4}{k}
\frac{(p_1^0-t_0 +m)(p_2^0-k_0-m)}
{[(p_1-t)^2-m^2][(p_2-k)^2-m^2]}  \nonumber \\ 
& & \quad \times \frac{1}{(\tv-\kv)^2 \kv^2
(\qv-\kv)^2 \tv^2 (\qv-\tv)^2} \, Q(\qv-\kv,\kv,\tv-\kv)  \, ,   
\plabel{Hgraph}
\ea
with
\beg
Q(\pv,\qv,\kv) := 
\pv\qv- \frac{(\pv \cdot \kv)(\qv \cdot \kv)} {\kv^2} \, .
\ee
In the centre-of-mass frame ${\bf P} = 0$, $p_1 = (P_0/2+p_0,\pv)$ 
and $p_2 = (-P_0/2+p_0,\pv)$. The integrations over $t_0$ and 
$k_0$ are quite simple, but the complicate angular structure 
seems to be a serious obstacle to the exact evaluation of the remaining 
integral. Observing that for the bound state the scales of the momenta
\footnote{This is the reason why a simple counting of vertices fails 
when estimating leading-order contributions of the Feynman graphs 
to the levels of bound states.} 
are $\pv, \qv \approx m \alpha$ and $p_0 \approx m \alpha^2$, 
it is natural to introduce rescalings as follows:
\begar 
p_0 &\to& \a^2 p_0  \, ,\nn \\
\pv &\to& \a \, \pv \, ,\nn\\
\qv &\to& \a \, \qv \, ,\plabel{scaling} \\
\tv &\to& \a \, \tv \, ,\nn \\
\kv &\to& \a \, \kv \, .\nn
\ea
As a consequence, the right-hand side of eq. \pref{Hgraph} 
is easily seen to retain a single power of $\alpha$ times  
an integral which is finite 
in the limit $\alpha \to 0$, remembering that the centre-of-mass energy 
is
$P_0 = 2m + O(\a^2)$. Therefore, the leading contribution in $\alpha$ 
of \pref{Hgraph} is simply obtained by putting $\a = 0$ in this integral. 
The resulting expression is 
\begar
H_{\ref{dfig5}.a} &=& 162 \, \pi^3 \a \int \ftt{3}{t} \ftt{3}{k} 
\frac{1}{k^2 (\qv-\kv)^2 t^2 (\qv-\tv)^2
(\tv-\kv)^2} \,  Q(\qv-\kv,\kv,\tv-\kv)  
\plabel{Hgraph2}
\ea
which only depends on $\qv$. The trick to simplify this expression is 
to write the inner products in terms of quadratic expressions that partly 
cancel the denominators:
\begar
Q(\qv-\kv, \kv, \tv-\kv) &=& \frac{1}{4} 
\Bigg[ \,2\,q^2 -k^2 -t^2 - (\qv-\kv)^2 + (\kv-\tv)^2 - (\qv-\tv)^2 
\nonumber \\
&-& \frac{t^2(\qv-\kv)^2}{(\kv-\tv)^2} +
\frac{t^2 (\qv-\tv)^2}{(\kv-\tv)^2} +
\frac{k^2 (\qv-\kv)^2}{(\kv-\tv)^2} -
\frac{k^2 (\qv-\tv)^2}{(\kv-\tv)^2} \Bigg] 
\, . \plabel{trick}
\ea
While the original integral \pref{Hgraph2} is finite, the integrals one
obtains from a single summand in \pref{trick} are not, so the 
evaluation of them requires some care. In the appendix we give the list 
of integrals necessary to evaluate \pref{Hgraph2}. The divergent integrals 
have been regularized with a cut-off in the integration region, but, 
as a check, we have verified that the same result can be obtained 
within  dimensional regularization. Removing the auxiliary regulators,
obviously all divergent terms cancel 
in the final expression, which can be written as
\beg
H_{\ref{dfig5}.a} 
= -\frac{81}{128}\pi (12- \pi^2) \frac{\a^3}{\qv^2} \, ,  
\plabel{H5a}
\ee
where we have restored the initial definition of $\qv$ (before scaling). 
Taking into account the factor $\a^3$ from the wave functions, 
we find, as a consequence of eq. \pref{H5a}, that the graph \ref{dfig5}a 
indeed yields contributions of 
$O(\a^4)$ to the energy levels. Therefore  
it must be included in an $O(\a^4)$ evaluation 
of the QCD potential $\CV$. Actually our result confirms some previous 
qualitative estimates \pcite{Duncan}--\pcite{Fein}.
\begin{figure}[htb]
\vskip 0.8truecm
\makebox[2truecm]{\phantom b}
\epsfxsize=12truecm
\epsffile{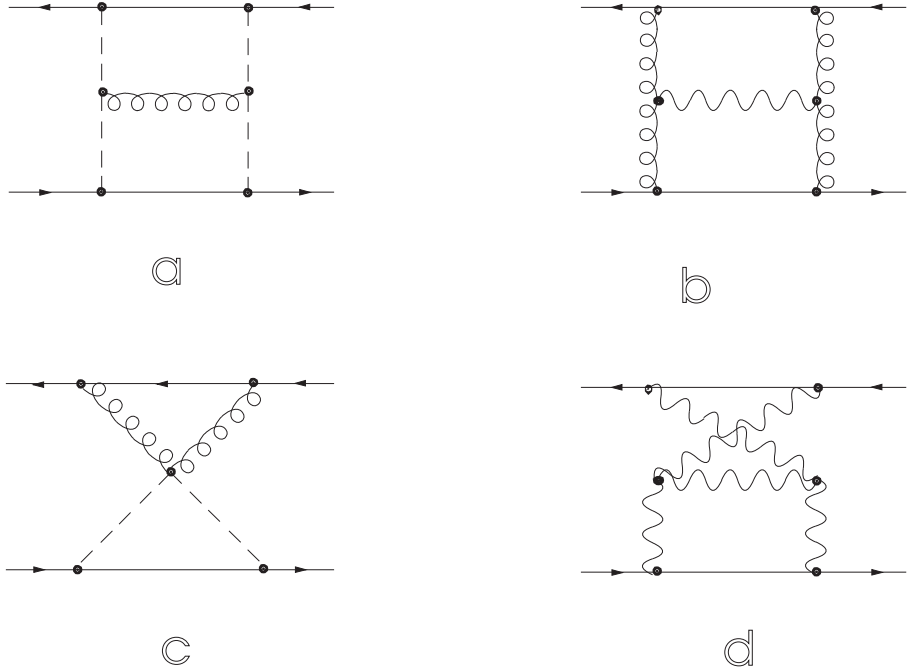}
\vskip 0.3truecm
\caption{{{\it Two-loop box graphs.}}}
\plabel{dfig5}
\vskip 0.8truecm
\end{figure}

Graph \ref{dfig5}b produces contributions to the energy levels that are 
of higher order than \ref{dfig5}a. In fact the 
insertion of a transverse gluon propagator between 
the wave function and the free quark propagator $S$ annihilates its 
leading contribution. But additional powers in the spatial 
momenta and consequently (see \pref{scaling}) in powers of $\a$ remain:
\begar 
S \left( \pm {P\over2} + p \right) &\approx& 
{i \over \pm p_0 + P_0/2 - E_p + i\epsilon}
\left( \lambda^{\pm} - {\pv \cdot \ggv \over 2 E_p} + \dots \right)  
+ \dots \, , \plabel{sf}
\\ 
\lambda^{\pm} \gamma_j S \left( \pm {P\over2} + p \right) &\approx& 
\lambda^{\pm} \gamma_j
{i \over \pm p_0 + P_0/2 - E_p + i\epsilon}
\left( - {\pv \cdot \ggv \over 2 E_p} \right)  + \dots \, , 
\ea
with $E_p := \sqrt{\pv^2 + m^2}$. However, we note that this graph contributes 
to the spin-dependent part of the interaction. 

Graph \ref{dfig5}c was considered in \pcite{KM1}, where it has 
been shown that the corresponding leading contribution to the energy 
level shift is $O(\a^5 \ln \a)$. 

Graphs like fig. \ref{dfig5}a and \ref{dfig5}b with crossed 
Coulomb lines (fig. \ref{dfig5}d) are irrelevant because they 
vanish due to group theoretical factors.

By connecting the quark lines between the interactions in the non-Abelian
vertex correction graph and graph \ref{dfig5}a, as shown in 
fig. \ref{dfig5a}, one obtains graphs which could be named non-Abelian 
Lamb shift graphs. Consider for instance the graph of fig. \ref{dfig5a}a  
with only one Coulomb gluon insertion, and call $t$ the momentum 
carried by the transverse gluon, $s$ and $k$ the momenta of 
the Coulomb gluon connecting it with the upper fermion line. 
Taking into account only the leading contribution to the fermion propagator 
(the first term of eq. \pref{sf}), we obtain
\begar
H^{(1)}_{\ref{dfig5a}.a} &=& -81 \, \pi^4 \a^4 \int \ftt{3}{t} \ftt{3}{k}
\ftt{3}{s} \frac{1}{t(\sv-\tv)^2 k^2 (\kv+\tv)^2 s^2 (\qv - \kv -\sv)^2} 
\nonumber\\
&~& \quad\quad\quad\times 
\frac{1}{(t+E_{p^\prime k} +E_{p^\prime k t} - P_0)
(t+E_{p s} +E_{p s t} - P_0)} \,  Q(\sv,\kv,\tv)  \, ,
\plabel{IIIgraph}
\ea
with $E_{p^\prime k} := \sqrt{(\pv^\prime - \kv)^2 + m^2}$, 
$E_{p^\prime k t} := \sqrt{(\pv^\prime - \kv - \tv)^2 + m^2}$, 
$E_{p s} := \sqrt{(\pv + \sv)^2 + m^2}$ and 
$E_{p s t} := \sqrt{(\pv + \sv - \tv)^2 + m^2}$. Scaling  all momenta 
according to eq. \pref{scaling}, $H^{(1)}_{\ref{dfig5a}.a}$  takes the form
\beg
H^{(1)}_{\ref{dfig5a}.a} 
= -81 \pi^4 \alpha^2 I(\alpha, \qv) \, ,  
\plabel{III2}
\ee
where $I$ is the integral of eq. \pref{IIIgraph} after scaling. 
Since it can be shown that the leading contribution of the derivative 
of $I$ with respect to $\alpha$ is proportional to $1/\alpha$ 
(scaling $t$ a second time and putting $\alpha = 0$ in the integral), 
the leading contribution of $H^{(1)}_{\ref{dfig5a}.a}$ is 
\beg
H^{(1)}_{\ref{dfig5a}.a} 
=  c_{\ref{dfig5a}.a}\alpha^2 \ln \alpha \, . 
\plabel{III3}
\ee From this argument it follows that all the graphs of fig. \ref{dfig5a} 
will contribute at least at the usual Lamb shift level $O(\a^5 \ln \a)$. 
\begin{figure}[htb]
\vskip 0.8truecm
\makebox[2truecm]{\phantom b}
\epsfxsize=12truecm
\epsffile{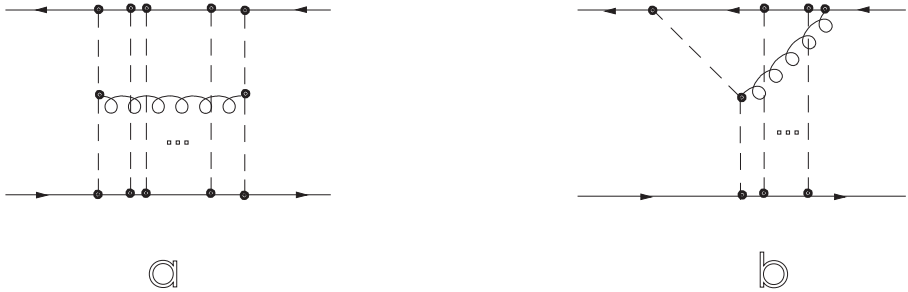}
\vskip 0.3truecm
\caption{{{\it Non-Abelian Lamb shift graphs.}}}
\plabel{dfig5a}
\vskip 0.8truecm
\end{figure}

\sect{QCD potential and renormalization schemes}

The $O(\a^4)$ correction from the non-Abelian box graph fig. \ref{dfig5}a 
leads to a slight enhancement of the attractive Coulomb 
force. This is the only contribution of this type in our
renormalization scheme, which is defined as follows: $m_{\rm t}$ is the 
pole mass, $Z_{\rm gluon}$ and $Z_{\rm 1F,Coulomb}$ are defined by a 
subtraction at $\m$ and 0, respectively. The updated potential given 
in \pcite{KM1} now reads 
\begar
 \CV &=& -\frac{\pv^4}{4 m^3} - \frac{\a \pi}{m^2} \d ({\bf r}) - \frac{\a }{2
m^2 r} \left( \pv^2 + \frac{{\bf r} ({\bf r}\cdot\pv)\pv} {r^2} \right) 
\nonumber\\ 
& & + \frac{3 \a }{2 m^2 r^3} {\bf L} {\bf S} +  \frac{\a }{2 m^2
r^3} \left( \frac{3  ({\bf r} {\bf S})^2} {r^2} - {\bf S}\,^2 \right) +
\frac{4 \pi \a }{3 m^2} {\bf S}^2 \d ({\bf r}) 
\nonumber \\ 
& & - \frac{33\a^2}{8 \pi r} ( \g + \ln \m r) +\frac{\a^2}{4\pi r}
\sum_{j=1}^{5} \left[\mbox{Ei}(-r m_j e^{\frac{5}{6}}) -\frac{5}{6} 
+ \2 \ln \left(\frac{\m^2}{m_j^2} + e^{\frac{5}{3}}\right) \right] 
+ \frac{9 \a^2}{8 m r^2} 
\nonumber \\ 
& &  - {81 \over 128}\pi(12 - \pi^2) 
\frac{\a^3}{4 \pi r} -  \frac{2 \a^3}{(16 \pi)^2 r} \left\{
27^2 \left[\frac{\pi^2}{6}+ 2(\g+\ln \m r)^2 \right] 
+ 576(\g+\ln \m r)\right\}   
\nonumber\\ 
& & - \frac{8}{9} \frac{4\pi \a_{\rm QED}(\m) \a}{r}- \sqrt{2} G_F m^2
\frac{e^{-m_H r}}{4 \pi r} + \sqrt{2} G_F m_Z^2 a_f^2 \frac{\d({\bf r})}
{m_Z^2} \left(7- \frac{11}{3} {\bf S}^2 \right)  
\nonumber \\ 
& & +\sqrt{2} G_F m_Z^2 a_f^2  \frac{e^{-m_Z r}}{2 \pi r}
\Bigg[1-\frac{v_f^2}{2a_f^2} - 
\left({\bf S}^2 - 3 \frac{({\bf S}{\bf r})^2}{r^2}\right)
\left(\frac{1}{m_Z r} + \frac{1}{m_Z^2 r^2}\right) 
\nonumber\\
& &\quad\quad\quad\quad\quad\quad\quad\quad\quad\quad\quad 
- \left({\bf S}^2 -\frac{({\bf S}{\bf r})^2}{r^2}\right)\Bigg] \, , 
\plabel{result} 
\ea
where $m_j$, $m_H$ and $m_Z$ are the quark, Higgs and $Z$ masses. 

However, the usual predictions for scattering measurements are in 
the $\overline{\mbox{MS}}$ scheme. Thus a short remark on the coupling 
constant to be used seems in order. It is customary to take the strong 
coupling constant $\as$ in the $\overline{\mbox{MS}}$ scheme. 
In particular experimental determinations are always given in terms of
$\as^{\overline{\rm MS}}(m_Z)$ \pcite{Altarelli}. 
But in bound-state calculations it is natural to use an $\as$ 
defined differently. This has been done in \pcite{KM1}. 
Clearly it should be possible to relate the two schemes.

The quark-antiquark potential in the $\overline{\mbox{MS}}$ scheme
with only light quarks to $O(\as)$ in the self-energy function 
of the Coulomb gluon can be taken for instance from \pcite{Billoire}. 
The contribution of heavy quarks in this scheme has been calculated 
in \pcite{Hagi}. We can thus write for the potential in momentum space
\beg 
\CV_{\overline{\rm MS}} = 
- \frac{4\pi \a}{\qv^2} \left[ 1 - \frac{\a}{4 \pi} \left(\frac{33}{4} 
\ln \frac{\qv^2}{\m^2} + \frac{31}{4} 
-  \sum_{i=1}^{n_f} \left[\frac{5}{6} -\frac{1}{2} 
\ln \left(\frac{\qv^2}{\m^2}+ \frac{m_j^2}{\m^2}e^{\frac{5}{3}}
\right) \right]  \right) \right] \, .
\ee
Comparison with the potential $\CV$ gives 
\beg 
\a^{\rm BS}=\a_{\overline{\rm MS}}
\left[1- \frac{\a_{\overline{\rm MS}}}{16 \pi}
\left( 31- \sum_{i=1}^{n_f} \left[ \frac{10}{3}
- 2 \ln \left(1+ \frac{m_j^2}{\m^2} e^{\frac{5}{3}}
\right) \right] \right) \right] \, ,
\plabel{alrel}
\ee
where BS denotes our bound-state scheme. It should be emphasized that
we have explicitly included the dependence on massive flavours in this
formula. Numerically, eq. \pref{alrel} means that our $\a$ is slightly 
smaller than the usual one, which is advantageous for our perturbative 
calculation. For $m_{\rm t} = 180 \,$ GeV, 
$\as^{\overline{\rm MS}}(m_Z) = 0.119\pm 0.006$, we obtain for a 
renormalization at $\m = 1/r_{\rm B}(\m)$ ($r_{\rm B}= 2/m\a$ 
is the Bohr radius of the bound state):
\begar
\m &=& 17.5 \pm 1 \, \mbox{GeV} \, , \\ 
\a^{\overline{\rm MS}}(\m) &=& 0.205 \pm 0.01 \, , \\ 
\a(\m)&=&0.19 \pm 0.01 \, .
\ea
Remember that the above values for $\a,\a^{\overline{\rm MS}}$ differ 
by a factor of  $4/3$ from $\as$.

For the comparison of the two schemes, the terms of the form $\as^n/\qv^2$ 
have been important. To compare to relative $O(\as^2)$, terms
up to $\as^3/\qv^2$ are needed in the potential. While the only term of 
this form in our scheme has been calculated in our present work 
(eq.\pref{H5a}), we are not aware of an analogous calculation of the terms of 
order $\as^2/\qv^2$ in the $\overline{\mbox{MS}}$ scheme. 
Therefore, at present the strong coupling 
constant in these two schemes can only be related to $O(\as)$. 

\sect{Conclusions}

In this note we present an analytical calculation of the contribution
of the QCD box graphs to the potential that would give rise to an $O(\a^4)$
energy level shift (eq. \pref{H5a}). This result is new and confirms previous
qualitative estimates (\pcite{Duncan}--\pcite{Fein}). 
It provides an example of a non-trivial contribution 
to $O(\a^4)$ of a two-loop graph, which certainly is not included e.g. in 
the running coupling constant and which is typical for a non-Abelian theory. 

While the calculated correction appears  
in our bound-state scheme as well as in the more common $\overline{\mbox{MS}}$ 
scheme, some corrections of the same form have not been calculated as yet to 
the required order in the latter. Thus a direct comparison of the two schemes 
is out of reach, at present. 

We also identified many non-Abelian graphs that give rise to $O(\a^5 \ln \a)$ 
contributions. These involve two-loop graphs as well as infinite 
chains of graphs, which could be called ``non-Abelian Lamb shift graphs".
\vspace{1cm}

{\bf Acknowledgement:} This work has been supported by the Austrian Science
Foundation (FWF), project P10063-PHY, within the framework of the
European Union Programme ``Human Capital and Mobility", Network 
``Physics at High Energy Colliders", contract CHRX-CT93-0357 (DG 12 COMA).

\appendix
\sect{Appendix}

In this appendix we give some of the integrals needed
to obtain \pref{H5a} from \pref{Hgraph2}.
The divergent integrals are regularized with cut-offs  
in the integration region. For the evaluation of these integrals 
within dimensional regularization we refer the reader to \cite {Muta}:
\begar
&~& \int \ftt{3}{t}{1\over t^2}{1\over (\qv-\tv)^2}
{1\over (\tv - \kv)^2} = {1\over 8}{1\over  q k}{1\over (\qv-\kv)^2} \, , \\
&~& \int \ftt{3}{t}{1\over (\qv-\tv)^2}
{1\over (\tv - \kv)^2} = {1\over 8 }{1\over |\qv-\kv|} \, , \\
&~&\int_{k>\epsilon_1 \, |q-k|>\epsilon_2} 
\ftt{3}{k}{1\over k^3}{1\over |\qv-\kv|^3}  = 
{1\over 2 \pi^2}{1\over  q^3}(2 \ln(q) - \ln(\epsilon_1) 
- \ln(\epsilon_2) - 1) \, , \\ 
&~&\int_{|q-k|>\epsilon_2} 
\ftt{3}{k}{1\over k^2}{1\over |\qv-\kv|^3}  = 
{1\over 2 \pi^2}{1\over q^2}(\ln(q) - \ln(\epsilon_2)) \, , \\ 
&~&\int_{k>\epsilon_1} 
\ftt{3}{k}{1\over k^3}{1\over |\qv-\kv|^2}  = 
{1\over 2 \pi^2}{1\over q^2}(\ln(q) - \ln(\epsilon_1)  + 1) \, , \\ 
&~&\int_{|q-k|>\epsilon_2} 
\ftt{3}{k}{1\over k}{1\over |\qv-\kv|^3}  = 
{1\over 2 \pi^2}{1\over q}(\ln(q) - \ln(\epsilon_2)) \, , \\ 
&~&\int_{k>\epsilon_1} 
\ftt{3}{k}{1\over k^3}{1\over |\qv-\kv|}  = 
{1\over 2 \pi^2}{1\over q}(\ln(q) - \ln(\epsilon_1) + 1) \, , \\ 
&~&\int_{k>\epsilon_1 \, |q-k|>\epsilon_2} 
\ftt{3}{k}{1\over k^3}{1\over |\qv-\kv|^3}  = 
{1\over 2 \pi^2}{1\over q^3}(2 \ln(q) - \ln(\epsilon_1) 
- \ln(\epsilon_2) - 1) \, . 
\ea
Finally a useful identity is
\beg 
\int \ftt{3}{k} \ftt{3}{t}
{1\over k^2}\left( {1\over (\qv-\kv)^2} - {1\over(\qv - \tv)^2} \right)
{1\over (\tv -\kv)^4}  = {1\over 32 \pi^2}{1\over q^2} \, .
\ee

\end{document}